\documentclass[reprint, amsmath, amssymb, aps, superscriptaddress, groupedaddress]{revtex4-2}
\usepackage{amsmath, braket, bm, amssymb, blindtext, epsfig, graphicx, array}
\usepackage{xcolor}
\usepackage{algorithm}
\usepackage{algpseudocode}

\usepackage{float, multirow}
\usepackage[colorlinks=true, allcolors=blue]{hyperref}

\usepackage[caption=false]{subfig}
\usepackage{hypcap} 

\newcommand\smallmath[2]{#1{\raisebox{\dimexpr \fontdimen 22 \textfont 2
      - \fontdimen 22 \scriptscriptfont 2 \relax}{$\scriptscriptstyle #2$}}}

\newcommand\smallotimes{\smallmath\mathbin\otimes}

\begin{document}

\title{Performance analysis of a filtering variational quantum algorithm}

\author{Gabriel Marin-Sanchez}
\email{gabriel.marin@quantinuum.com}
\affiliation{Quantinuum, Partnership House, Carlisle Place, London SW1P 1BX, United Kingdom}
\author{David Amaro}
\email{david.amaro@quantinuum.com}
\affiliation{Quantinuum, Partnership House, Carlisle Place, London SW1P 1BX, United Kingdom}

\date{\today}

\begin{abstract}
Even a minor boost in solving combinatorial optimization problems can greatly benefit multiple industries. Quantum computers, with their unique information processing capabilities, hold promise for delivering such enhancements. The Filtering Variational Quantum Eigensolver (F-VQE) is a variational hybrid quantum algorithm designed to solve combinatorial optimization problems on existing quantum computers with limited qubit number, connectivity, and fidelity.  In this work we employ Instantaneous Quantum Polynomial circuits as our parameterized quantum circuits. We propose a hardware-efficient implementation that respects limited qubit connectivity and show that they halve the number of circuits necessary to evaluate the gradient with the parameter-shift rule. To assess the potential of this protocol in the context of combinatorial optimization, we conduct extensive numerical analysis. We compare the performance against three classical baseline algorithms on weighted MaxCut and the Asymmetric Traveling Salesperson Problem (ATSP). We employ noiseless simulators for problems encoded on 13 to 29 qubits, and up to 37 qubits on the IBMQ real quantum devices. The ATSP encoding employed reduces the number of qubits and avoids the need of constraints compared to the standard QUBO / Ising model. Despite some observed positive signs, we conclude that significant development is necessary for a practical advantage with F-VQE.
\end{abstract}

\maketitle
\section{Introduction}
Combinatorial optimization, known for its inherent computational complexity, consists in finding the best solution among a set of discrete choices~\cite{comb-opt}. It finds applications in diverse fields ranging from logistics~\cite{Vogiatzis2013} and supply chain~\cite{van2005modeling} to healthcare~\cite{Kim2013} and finance~\cite{PUERTO2022105701}, so even a small improvement at solving these problems can have a huge impact in multiple industries.

In practical applications, a basic exhaustive search for the optimal solution is often computationally infeasible, given the vast search space and inherent complexity of the problem. Paradigms like Branch and Bound~\cite{bnb} and Dynamic Programming~\cite{P-550} leverage the structure of the problem to implement a clever exhaustive search, considerably speeding up the search for an optimal solution. Alternatively, heuristic-based techniques offer practical and often efficient approaches to finding reasonably good solutions~\cite{RABBOUCH2023407, heuristics}.
The development of cutting-edge approaches, such as metaheuristics~\cite{metaheuristics, metaheuristics2} and machine learning~\cite{BENGIO2021405}, has led to impressive advancements in solving combinatorial problems.

Nonetheless, quantum computing is being explored as a fundamentally different approach to combinatorial optimization~\cite{2312.02279}. However, while quantum algorithms for searching the solution like Grover's~\cite{grover} exhibit theoretical quadratic advantages, practical applications are impeded by the overhead from Quantum Error Correction~\cite{PRXQuantum.2.010103} when challenges such as noise and errors in quantum environments~\cite{Kaderi2022PerformanceOU} are considered.

Variational Quantum Algorithms (VQA) minimize these challenges by using minimal parameterized quantum circuits. Through iterative adjustments, they approximate operations, adapt to diverse hardware and noise patterns~\cite{cerezo2021a,Benedetti_2019}. Despite being initially introduced as an intermediate solution for Noisy Intermediate Scale Quantum  (NISQ) devices, their simplicity and versatility make them an interesting tool for fault-tolerant quantum computers as well, since the user can control the size and structure of the circuits~\cite{Sayginel2023AFV}. 

One promising example of VQA for combinatorial optimization is the Filtering-VQE (F-VQE)~\cite{Amaro_2022, fvqe2}, which showed improved performance compared to other quantum algorithms like the Variational Quantum Eigensolver (VQE)~\cite{vqe} and the Quantum Approximate Optimization Algorithm (QAOA)~\cite{farhi2014}. It works by iteratively boosting the probability of sampling the best solutions from a quantum state. At each step F-VQE trains a parameterized quantum circuit to approximate the action of a filtering operator on the quantum state obtained at the previous step, hence increasing the probabilities of sampling good solutions and decreasing the probabilities for sampling the bad solutions. At the end of the algorithm F-VQE outputs a list of the best solutions sampled.

In this article, we expand the F-VQE algorithm by using Instantaneous Quantum Polynomial (IQP) circuits as parameterized quantum circuits. These circuits check various necessary boxes for VQAs: sampling from them is conjectured to be classically hard~\cite{Bremner2010, Bremner2015}, even in the presence of some noise~\cite{Bremner2016}, and are not expected to contribute to the presence of barren plateaus, due to their limited expressibility~\cite{Larocca2022, Ragone2023, Bowles2023}. Additionally, they present some advantageous features. As we show in this work, IQP circuits can be implemented with a hardware-efficient circuit that respects the limited qubit connectivity, and halve the number of circuit executions to compute the gradient with the parameter-shift rule.

In this work we perform extensive numerical analysis of the F-VQE performance on two well-known optimization problems that have captivated researchers across multiple disciplines: the weighted MaxCut and the Asymmetric Travelling Salesperson Problem (ATSP). In particular, we run up to 256 random instances for qubit sizes from 13 to 29 using noiseless state-vector simulations. Instead of the standard Quadratic Unconstrained Binary Optimization (QUBO) form~\cite{10.3389/fphy.2014.00005} or equivalently, the Ising model, we encode the ATSP as a generalized cost function that reduces the number of qubits and avoids the need of imposing additional constraints~\cite{glos2022}. 

The large amount of data generated serves us to study the effects of barren plateaus~\cite{bp1} and the performance of F-VQE against three classical baseline algorithms: brute-force search (BFS), simulated annealing (SA), and a purely classical version of F-VQE that we introduce in this work. BFS serves as a basic health check, SA represents one of the most simple, yet flexible and high-performing classical algorithms~\cite{saHenderson}, while the classical F-VQE serves us to assess the relevance of purely quantum effects. We additionally study the performance drop due to noise by solving weighted MaxCut and ATSP instances of up to 37 qubits on IBMQ devices~\cite{ibmq} with F-VQE.

The ultimate goal of quantum computing is to show an advantage against its classical counterparts, so building a comparison against diverse classical algorithms is a pivotal step to understanding the potential of F-VQE. In this work we observe positive signs of the performance of F-VQE against three baseline classical algorithms in the regime of noiseless circuits of the size we can simulate and using the number of cost function evaluations as a proxy of real runtime. However, we conclude that significant development will be necessary to provide an advantage against the state-of-the-art classical algorithms in a realistic regime. 

This article is structured as follows. Section~\ref{sec:fvqe} introduces F-VQE. In Section~\ref{sec:iqp} we present the IQP and classical version of F-VQE. Section~\ref{sec:generalised} discuses the design principles of the generalized cost function and applies them to the encoding of the ATSP. We exhibit the numerical and experimental results in Section~\ref{sec:results} exhibits. Finally, we summarize this article and discuss the conclusions in Section~\ref{sec:conclusions}.

\section{Filtering Variational Quantum Eigensolver (F-VQE)}\label{sec:fvqe}
An optimization problem is defined as the task of finding candidate solutions--usually represented by a string of $N$ binary values $\bm{x}\in\{0,1\}^N$--that minimize the cost function $C(\bm{x})$ of the problem. Measuring a $N$-qubit quantum state $\ket{\psi}$ in the computational basis $\ket{\bm{x}}=\bigotimes_{q\in[N]}\ket{x_q}$ returns a sample candidate solution from the distribution $P_{\psi}(\bm{x}) = |\braket{\bm{x}|\psi}|^2$. In this work we employ the convenient notation $[a]\equiv\{1, 2, \ldots, a\}$ for every integer $a>0$. The goal of F-VQE and most VQAs is to train a $N$-qubit parameterized quantum state $\ket{\psi(\bm{\theta})}$ with $M$ parameters $\bm{\theta}\in[-\pi/2,\pi/2]^M$ to produce a probability distribution $P_{\psi(\bm{\theta})}(\bm{x})$ with high probability of sampling good (low cost) solutions. In Section~\ref{sec:iqp} we introduce IQP circuits as our parameterized quantum ansatz.

In the particular case of F-VQE~\cite{Amaro_2022}, the training consists in, at every step of the algorithm, approximating the action of a filtering function $f(c;\tau)$, i.e., a positive and monotonically decreasing function in the cost $c\geq 0$, on the probability distribution obtained at previous step. The exactly filtered distribution 
\begin{equation}
    P^{(f)}_{\psi}(\bm{x}) = f^2(C(\bm{x});\tau)P_\psi(\bm{x}) \mathbb{E}^{-1}_\psi(f^2)
\end{equation}
achieves a larger probability of sampling good solutions and a lower probability of sampling bad (high cost) solutions. Here, the expected value $\mathbb{E}_\psi(f^2)$ of the filtering function is the normalization constant. A large filtering strength $\tau>0$ intensifies this effect. In this work we employ the inverse filter $f(c;\tau)=c^{-\tau}$, as it was identified as the best-performing filtering function in~\cite{Amaro_2022}. 

The training process is the following. F-VQE starts from the uniform distribution of all candidate solutions $P_{\psi_0}(\bm{x}) = 2^{-N}$, generated by an initial quantum state $\ket{\psi_0} \equiv \ket{\psi(\bm{\theta}_0)}$. Suitable parameters $\bm{\theta}_0$ to produce this state can easily be found in most common parameterized quantum ans\"atze. At each step $t\in[T]$ F-VQE aims to minimize the loss function given by the infidelity  
\begin{equation}\label{eq:loss}
    \mathcal{L}_t(\bm{\theta}) = 1-\left(\sum_{\bm{x}\in\{0,1\}^N} \sqrt{P_{\psi(\bm{\theta})}(\bm{x})P^{(f)}_{\psi_{t-1}}(\bm{x})}\right)^2
\end{equation}
between the distribution produced by the parameterized ansatz $\ket{\psi(\bm{\theta})}$ and the exact filtered distribution from the state $\ket{\psi_{t-1}}\equiv\ket{\psi(\bm{\theta}_{t-1})}$ obtained at the previous step. Notice that, differently to the original F-VQE work~\cite{Amaro_2022}, we have defined the loss function in terms of probability distributions, rather than in terms of the quantum states that produce them. This definition does not change the algorithm and will become useful when we introduce the classical version of F-VQE. 

For those parameterized quantum ans\"atze where each parameter $\theta_k$ is present in only one rotation gate of the form $R_{\sigma_k}(\theta_k) = \exp(-i \theta_k \sigma_k/2)$ along the direction of a non-trivial Pauli operator $\sigma_k\in\{I, X, Y, Z\}^{\smallotimes  N}\setminus I^{\smallotimes  N}$, the gradient (up to a multiplicative positive constant) can be calculated via the parameter-shift rule~\cite{Mitarai2018, Schuld2019}. In Appendix~\ref{sec:app-grads} we derive the resulting gradient:
\begin{eqnarray}
    \left. \dfrac{\partial \mathcal{L}_t(\bm{\theta})}{\partial \theta_k} \right|_{\bm{\theta}_{t-1}} &&\propto \mathbb{E}_{\psi_{t-1}^{-k}}(f) - \mathbb{E}_{\psi_{t-1}^{+k}}(f), \label{eq:eq_exa_gra} \label{eq:grad-prob}\\
    \ket{\psi_{t-1}^{\pm k}} &&= \ket{\psi(\bm{\theta}_{t-1}\pm\pi\bm{e}_k/2)}. \label{eq:shifted}
\end{eqnarray}
These quantum states are obtained by shifting the $k$-th parameter at step $t-1$ by a quantity of $\pm \pi/2$. The gradient in Eq.~\eqref{eq:eq_exa_gra} is estimated (up to a multiplicative positive constant) by taking $s$ samples $S_t^{\pm k}$ from these two quantum states:
\begin{equation} \label{eq_app_gra}
    g_{tk} = \frac{1}{s}\sum_{\bm{x}\in S_t^{-k}}f(C(\bm{x});\tau) - \frac{1}{s}\sum_{\bm{x}\in S_t^{+k}}f(C(\bm{x});\tau).
\end{equation}
Finally, the parameters are updated at every step from a single normalized gradient descent step as $\bm{\theta}_{t} = \bm{\theta}_{t-1} - \eta\bm{g}_t/\lVert \bm{g}_t \rVert $, where $\eta$ is the learning rate. Notice that the positive constant disappears under the normalization of the gradient. As a side note, for $\tau=-1$, the filter function equals to the cost function, recovering the gradient of VQE with the parameter-shift rule.

\begin{figure}[t!]
  \centering
  \includegraphics[width=1\columnwidth]{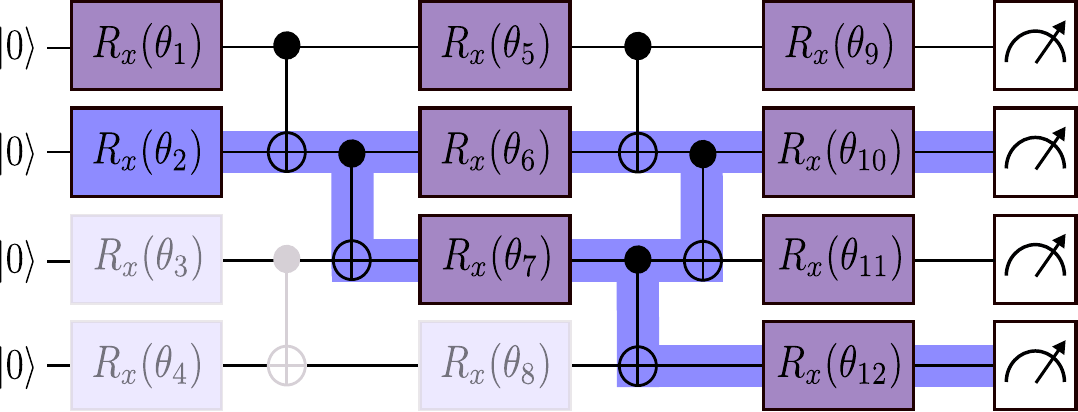}
  \caption{Illustration of the ansatz used to obtain an IQP circuit with $\ell=3$. Highlighted in blue is the propagation process of parameter $\theta_2$. When encountering the control part of the CNOT, the rotation propagates to the targeted qubit, cancelling the operation if such qubit was already affected. At the end of the circuit, this parameter acts on qubits 2 and 4 with the unitary $\exp{(-i\theta_2X_2X_4/2)}$. In the same way, notice that some parameters have the same contributions, like $\theta_4$, $\theta_8$ and $\theta_{12}$, and $\theta_3$ and $\theta_{11}$. Therefore we can delete the repeated gates starting from the left. The CNOT operations that act trivially after the deletion process are also deleted.}
  \label{fig:ansatz}
\end{figure}

\section{IQP circuits as parameterized quantum circuits}\label{sec:iqp}
Sampling from the family of Instantaneous Quantum Polynomial (IQP) circuits is conjectured to be classically hard~\cite{Bremner2010, Bremner2015}, even in the presence of some noise~\cite{Bremner2016}. Moreover, sampling from sparse IQP circuits in a quantum computer involves a minimal circuit depth and number of gates from the native gate set of most quantum devices. This made IQP circuits an attractive proposal for an early demonstration of quantum supremacy~\cite{lundQuantumSamplingProblems2017,Paletta2024}. In this work we propose them as the parameterized quantum ansatz for F-VQE.
A general parameterized IQP ansatz applies parameterized Pauli-$X$ rotations on $M$ subsets $Q_k\subset[N]$ of qubits as
\begin{eqnarray} \label{eq_iqp_gen}
    \ket{\psi(\bm{\theta})} &&= \exp\left(-i \sum_{k\in[M]} \frac{\theta_k}{2} X_{Q_k} \right)\ket{0}^{\smallotimes  N}, \\
    X_{Q_k} &&= \bigotimes_{q\in Q_k} X_q.
\end{eqnarray}

In this section we show that only one quantum circuit per parameter needs to be sampled to evaluate the approximate gradient in Eq.~\eqref{eq_app_gra} and how generic IQP circuits can be implemented as hardware-efficient circuits. Finally, we introduce a fully classical ansatz that replicates the IQP structure. F-VQE trained on this ansatz provides a fully classical version of F-VQE that allows us to asses the relevance of quantum effects. 

\subsection{Parameter-shift rule with a single circuit}
Aside for the minimal implementation requirements to obtain quantum circuit that is classically hard to simulate, we choose the IQP ansatz because it only needs to sample one quantum circuit to evaluate each partial derivative with the parameter-shift rule. Most ans\"atze require two or more~\cite{Wierichs2022generalparameter}. However, the total number of times that the cost function must be evaluated remains the same.

For the IQP ansatz, the two shifted states in Eq.~\eqref{eq:shifted} differ (up to a global phase) on the application of a bit-flip $X_{Q_k}$. 
Consequently the probability of sampling any binary string $\bm{x}$ in one shifted circuit coincides with the probability of sampling the binary string $\bm{x} \oplus \bm{q}_k$ obtained by flipping the positions corresponding to the qubits in $Q_k$. Here $\bm{q}_k$ is a binary string with $1$s only on the positions corresponding to the qubits in $Q_k$. Using this result, for each parameter and step we implement and sample only the $+$-shifted circuit and evaluate the gradient in Eq.~\eqref{eq_app_gra} as
\begin{equation}\label{eq:grad}
    g_{tk} = \frac{1}{s}\sum_{\bm{x}\in S_t^{+k}}f(C(\bm{x}\oplus \bm{q}_k);\tau) - f(C(\bm{x});\tau).
\end{equation}

\subsection{Hardware-efficient circuit implementation}
In this section we show that very complex IQP circuits can be implemented on hardware with a limited qubit connectivity without the need of using SWAP gates. This provides a faster implementation subject to less noise sources. 

As depicted in Fig.~\ref{fig:ansatz}, our implementation of the IQP ansatz consists in layers of single-qubit parameterized $X$-rotations $R_x(\theta_k)= \exp(-i \theta_k X/2)$ and CNOT gates between nearest neighbors in a one-dimensional chain. This ansatz produces particular instances of the IQP circuit as appearing in Eq.~\eqref{eq_iqp_gen}. As illustrated in the figure, one can propagate every single-qubit rotation through the product of all CNOT gates towards the end of the circuit following the propagation rules: $\mathrm{CNOT}_{ij}X_i = X_i X_j \mathrm{CNOT}_{ij}$ and $\mathrm{CNOT}_{ij}X_j = X_j \mathrm{CNOT}_{ij}$, where $i$ ($j$) is the control (target) qubit of the CNOT gate. Now, CNOT gates act trivially on the $\ket{0}^{\smallotimes  N}$, while the remaining $X$-rotations take support on subsets $Q_k$. These subsets can be obtained by simple inspection, as described in the figure. The number $\ell$ of layers in the noiseless simulations is chosen so that every pair of qubits is part of at least one subset $Q_k$. 
 
This requires the number of layers to grow linearly with the number of qubits.   

Notice that certain parameters may result in repeated qubit subsets, as shown in Fig.~\ref{fig:ansatz}. Consequently, some operations can be removed to simplify the circuit. The initial quantum state $\ket{\psi_0}$ of F-VQE is obtained by setting the parameters in the first $\ell-1$ layers to $0$ and the parameters in the last layer to $\pi/2$. In the IBMQ devices where a sufficiently long chain can not be found we adapt the circuit connectivity to that of the device to avoid the use of SWAP gates that introduce additional noise. We provide an example in Appendix~\ref{sec:app-ibmq}.

\subsection{Classical ansatz} \label{subsec:class_ans}
We introduce a purely classical ansatz that mirrors the structure of the IQP circuit but lacks any quantum properties. Following a similar approach to that in~\cite{weitz2023subuniversal}, the objective of this classical algorithm is not to simulate the IQP ansatz, but rather to determine whether quantum effects give any advantage to the performance of F-VQE. This ansatz consists in a parameterized probabilistic circuit that applies bit-flip channels on the same qubit subsets $Q_k$ of Eq.~\eqref{eq_iqp_gen}. We use the density matrix formulation but notice that this ansatz represents a fully classical probability distribution:
\begin{eqnarray}
    \rho(\bm{\theta}) &&= \mathcal{E}_{\theta_M}\circ\dots\circ\mathcal{E}_{\theta_1} \left(\ket{0}\bra{0}^{\smallotimes  N}\right), \label{eq:class_ans}\\
    \mathcal{E}_{\theta_k} (\rho) &&= \cos^2(\theta_k/2)\rho+\sin^2(\theta_k/2) X_{Q_k} \rho X_{Q_k}. \label{eq:bitflip}
\end{eqnarray}

As an illustration, if we have the unitary rotation $\exp(-i\theta_2 X_2X_4/2)$ acting on qubits 2 and 4 in the IQP ansatz, in the classical ansatz we have a bit-flip channel that flips bits 2 and 4 with probability $\sin^2(\theta_2/2)$. As we show in Appendix~\ref{app:dephasing}, the classical ansatz is equivalent to the original IQP ansatz with fully dephasing channels~\cite{Nielsen_Chuang_2010}, hence destroying any coherence, acting after each layer.

As in the quantum setting, our classical ansatz produces a probability distribution $P_\rho(\bm{\theta}) = \braket{\bm{x}|\rho|\bm{x}}$ over candidate solutions. Hence, the same training algorithm of F-VQE can be applied with the same aim of increasing the probability of sampling good solutions. Every step described in sections~\ref{sec:fvqe} and~\ref{sec:iqp} can be replicated for the classical ansatz if the pure quantum state $\ket{\psi}$ is replaced by the classical state $\rho$. In Appendix~\ref{sec:app-grads} we show that both ans\"atze have the same gradients and both benefit from the reduction in circuit executions.
\section{QUBO and generalized cost function} \label{sec:generalised}
The typical approach to solving combinatorial optimization problems with VQAs starts by formulating the problem in the Quadratic Unconstrained Binary Optimization (QUBO) form. In this form, the cost function is quadratic in the binary variables that define candidate solutions. The limitation to quadratic terms is convenient traditionally in the context of quantum optimization because such cost function maps naturally to a two-body classical Ising Hamiltonian, which is an ideally suited object for a quantum computer. For example, quantum annealing~\cite{Kadowaki1998} and QAOA only need two-body interactions to evolve the quantum state under this Hamiltonian. However the QUBO form is very natural for some problems like MaxCut, but not so much for the ATSP. In light of this limitation, we use a generalized cost function for ATSP that is more natural than the QUBO form. This concept has been proposed in previous studies~\cite{glos2022,Amaro_2022} and recently tested on F-VQE for job shop scheduling problems~\cite{Schmid2024highly}. However, its performance on VQAs has not been extensively explored. 

\subsection{QUBO for weighted MaxCut}
The weighted MaxCut involves partitioning the vertices of an undirected weighted graph into two sets such that the total weight of the edges cut by the partition is maximized. This optimization problem has applications in fields such as graph theory~\cite{poljak1995maximum} and statistical physics~\cite{barahona1988}. It has been proven to be NP-hard if a solution sufficiently close to optimality is desired, but classical heuristics can reach solutions with large approximation ratios in polynomial time~\cite{goemans-williamson}. Due to its well-established classical performance and natural mapping to a QUBO, it commonly serves as a benchmark for evaluating quantum algorithms like QAOA~\cite{farhi2014}.

Consider a 3-regular undirected graph with a set $[n]$ of $n$ vertices and a set $E$ of edges $e\in E$ with random positive weights $w_e\in(0,1]$. A cut is represented by $n$ binary variables such that $x_v=1(0)$ if node $v\in[n]$ is in one set (or in the complementary). Notice that swapping the tags $0,1$ for the two sets preserves the problem definition. We break this symmetry by fixing the tag of the last vertex to 0, hence reducing the number of variables to $N=n-1$. The cost function associated to the weighted MaxCut problem can be defined as
\begin{align}
    C(\bm{x})=&-\sum_{\substack{e=\{v,n\}\in E}}w_e x_v \nonumber\\&- \sum_{\substack{e=\{u,v\ne n\}\in E}}w_e( x_u+x_v-2x_ux_v).
\end{align}

\subsection{Generalized cost function for ATSP} \label{subsec:atsp}
The ATSP generalizes the classical travelling salesperson problem (TSP) to include asymmetric city-to-city distances, a realistic scenario with several applications in vehicle routing and job-shop schedulling problems~\cite{lenstra1975, matai2010}. The solution of the problem is the shortest route that visits every of the $n$ cities once and returns to the initial city. There are theoretical inapproximability results that show the NP-hard complexity to approximate the ATSP within a certain ratio~\cite{papadimitriou}. 

The QUBO form for the ATSP requires $n^2$ binary variables with values $x_{ij}=1(0)$ if the $i$-th visited city is (is not) city $j$~\cite{10.3389/fphy.2014.00005}. For quantum computers, such rapid quadratic growth in the number of qubits becomes a challenge. Additionally, since most possible binary strings do not represent feasible routes, constraints must be imposed on the binary variables. In the QUBO form they are imposed as large penalty terms in the cost function, increasing the resolution necessary to distinguish feasible solutions~\cite{Ayodele2022} and adding more parameters that require calibration~\cite{Salehi2022}.
\\

Fortunately, for algorithms like VQE~\cite{Garcia-Saez:2018ezk} or F-VQE, where the ansatz is independent of the problem and samples from the quantum computer represent candidate solutions, the limitations of the QUBO form are unnecessary. In the case of F-VQE, as long as the cost $C(\bm{x})$, which may be called generalized cost function, can be efficiently evaluated classically for every candidate solution, the gradient in Eq.~\eqref{eq:grad} can be computed normally. Notice that this generalized cost function can be extremely complex, running several subroutines, loops, load external data, compute continuous variables, etc... For example, this flexibility may provide access to databases for machine learning tasks like feature selection~\cite{Zoufal2023variationalquantum}. Furthermore, this adaptability is convenient to obtain a problem formulation that requires a reduced number of binary variables by, for instance, encoding the candidate solutions in a way more natural to the optimization problem at hand. 

Instead of the QUBO form, in this work we employ the generalized cost function described in Algorithm~\ref{alg:atsp} for ATSP. Here candidate solutions are more naturally represented by permutations of the cities. Since every permutation is a valid route that visits every city once, there is no need for constraints, nor for the associated penalty terms. Besides, only $N\sim n\log n$ qubits are required. In~\cite{glos2022} a generalized cost function for TSP was proposed but no experiments were performed. We extend this cost function to the ATSP and study the performance extensively.

The input of the generalized cost function is a string $\bm{x}$ of $N$ binary values representing a candidate solution, on which three subroutines are executed to return the associated cost $C(\bm{x})$. The first subroutine is a binary or gray code to represent $\bm{x}$ as an integer $x\in\{0, 1, \ldots, 2^N-1\}$. The second subroutine is the Lehmer code~\cite{Lehmer1960TeachingCT}, which, given $x$, generates the associated permutation $\bm{\sigma}$ of the first $n-1$ cities. A route $\bm{r}$ is obtained by appending the $n$-th city to the permutation. This trick saves some qubits by breaking the rotational symmetry of the problem when forcing the $n$-th city to be the last visited city. Third, the cost of the route can be read directly from $\bm{\sigma}$ and the costs $W_{ij}$ to go from city $i$ to city $j$. Finally, the total cost of the route is returned as the output of the generalized cost function.

\begin{algorithm}[H]
\caption{Generalized cost function for the ATSP.}\label{alg:atsp}
\textbf{Input:} List of $n$ cities $\bm{c} = (c_1, c_2, \ldots, c_n)$, adjacency matrix $W$ and a binary string $\bm{x}$ with $N=\lceil \log_2((n-1)!) \rceil$ bits \\
\textbf{Output:} Route distance $C(\bm{x})$
\begin{algorithmic}[1]
\State $x \gets \bm{x}$  \Comment{Binary or any gray code}
\State $\bm{\sigma} \gets x, \, \bm{c}$ \Comment{Lehmer code}
\State $\bm{r} \gets \bm{\sigma}, \, c_n$ \Comment{append last city}
\State $C(\bm{x}) = W_{\sigma_n, \sigma_1} + \sum_{i=1}^{n-1}W_{\sigma_i, \sigma_{i+1}} \gets \bm{r}, \, W$
\end{algorithmic}
\end{algorithm}
The number of permutations of $n-1$ cities is $(n-1)!$, so this generalized cost function needs exactly $N=\lceil \log_2((n-1)!) \rceil \sim n \log n$ qubits. There is an excess of $2^N - (n-1)!$ binary strings, that simply produce double-degeneracy of some candidate solutions. 

\section{Results}\label{sec:results}
We test the performance of F-VQE with the IQP ansatz against various algorithms on noiseless simulations and IBMQ devices. For the noiseless simulations we additionally study the effect of barren plateaus. We start this section with a description of the experimental set-up and then continue to the results from the experiments. 

\subsection{Set-up}
In this work we test the performance of the following algorithms.

\textit{F-VQE} with the IQP ansatz as described in Sections~\ref{sec:fvqe} and~\ref{sec:iqp}. We set the total number of training steps to $T=200$. The remaining hyperparameters are the number $s$ of samples per circuit, the learning rate $\eta$ and the filtering strength $\tau$. To select the best hyperparameters we test four different settings, with high and low $s$, and with constant and variable $\eta$ and $\tau$. As shown in Appendix~\ref{sec:app-hyp} the four settings provide similar performance. We choose the setting with high $s$ and constant $\eta$ and $\tau$ because it solves a slightly larger fraction of problem instances.

\textit{F-VQE with the classical ansatz} as described in Section~\ref{subsec:class_ans}. We perform the same analysis of hyperparameters and choose the same settings which also solve the largest fraction of problem instances. We are interested in this classical ansatz because, given the close relation with the IQP ansatz, any difference in performance might be an indication of the role played by the quantum effects. 

\textit{VQE with the IQP ansatz}. As mentioned in Section~\ref{sec:fvqe}, the gradient of VQE can be recovered by setting $\tau=-1$ in F-VQE. We study the effects of barren plateaus for both quantum algorithms to assess the effect of the filtering function. The same number of samples is used, but with a different learning rate $\eta=0.35$, that was found to perform better. 

\textit{Brute-force search (BFS)}. Random candidate solutions are drawn uniformly from the space of possible solutions without repetition. As for the other algorithms, we can report the best approximation ratio obtained from any number of samples. As a bare minimum to show promising performance, every good quantum algorithm for optimization should clearly outperform brute-force search.   

\textit{Simulated Annealing (SA)} is one of the most successful heuristic algorithms for combinatorial optimization~\cite{saHenderson} due to its simplicity and wide range of applicability. This method samples candidate solutions from an approximate thermal distribution generated by an adaptation of the Metropolis-Hastings algorithm with a scheduled temperature reduction. The simple version considered here should be considered as another baseline algorithm that every good quantum algorithm for optimization should beat. For MaxCut, the transition operation of SA is flipping one random bit, while for ATSP, we swap two adjacent cities selected randomly. As recommended in~\cite{ameur2004} we set the initial temperature parameter to 5, so that the probability of accepting the transition from the solution with maximum cost value to the solution minimum cost is 0.8. To select the final temperature, we test the values of 1, 0.1, 0.01 and 0.001 and run all problems with a given qubit size. We use the value that solves more problems optimally. \\ 

\textit{Approximation ratio} as a benchmark metric. 
Given a cost function $C(\bm{x})$, with minimum and maximum values $C_\mathrm{min}$ and $C_\mathrm{max}$, the approximation ratio is defined as
\begin{equation}
    A(\bm{x}) = \frac{C_\mathrm{max} - C(\bm{x})}{C_\mathrm{max} - C_\mathrm{min}}. 
\end{equation}
This metric measures the quality of the solutions and returns values between 0 and 1, corresponding to the worst and best solution qualities, respectively. We can compute the exact minimum and maximum values for the problem instances generated in this work via brute-force search. 
We consider the realistic scenario where only the best solution ever sampled during the entire execution of the algorithm is returned. At each step of a algorithm, we report the approximation ratio as the \textit{best approximation ratio sampled} from the start of the algorithm until that step. \\

\textit{Generation of problem instances.} For the noiseless study, we generate 256 random problem instances of each problem class (weighted MaxCut and ATSP) for problem sizes of up to 26 qubits, 128 instances for 27 qubits and 32 instances for the 29 qubits case. In addition, we also generate 3 instances of 19, 25, 29, 31, 33 and 37 qubits for weighted MaxCut and 19, 22, 26, 29, 33 and 37 qubits for ATSP for the experiments on IBMQ devices. 

In Appendix~\ref{sec:app-pg} we show that, for all problem instances generated, the number of solutions with an approximation ratio larger than $a$ decays exponentially in $a$. This is a sanity check to avoid the situation where good solutions are obtained simply due to the existence of a large fraction of good solutions. 

As described in~\cite{Amaro_2022}, F-VQE re-scales the problem's cost function into the range $(0,1]$. To bring the cost function into such range we need to compute upper and lower bounds to the maximum and minimum costs, respectively. For weighted Maxcut, the upper bound can be trivially set to 0, and the lower bound is computed using semidefinite programming (SDP)~\cite{Helmberg2000}. For ATSP we employ the Chu-Liu-Edmonds' algorithm~\cite{chu-liu,edmonds1967optimum} to find the optimum branching of each city and compute both bounds. Computing these bounds for F-VQE can always be done efficiently with SDP.\\

\textit{Quantum software.} Circuit design is performed by TKET~\cite{Sivarajah_2021}. Noiseless state-vector simulations are performed by Qulacs~\cite{Suzuki2021qulacsfast}. Experiments in IBMQ devices are performed via Qiskit Runtime~\cite{ibmq}. For the analysis of barren plateaus we use qujax~\cite{Duffield2023} to compute the exact gradients of the circuits. We provide a repository~\cite{github-data} with all the collected data.


\begin{figure*}
  \centering
  \includegraphics[width=2\columnwidth]{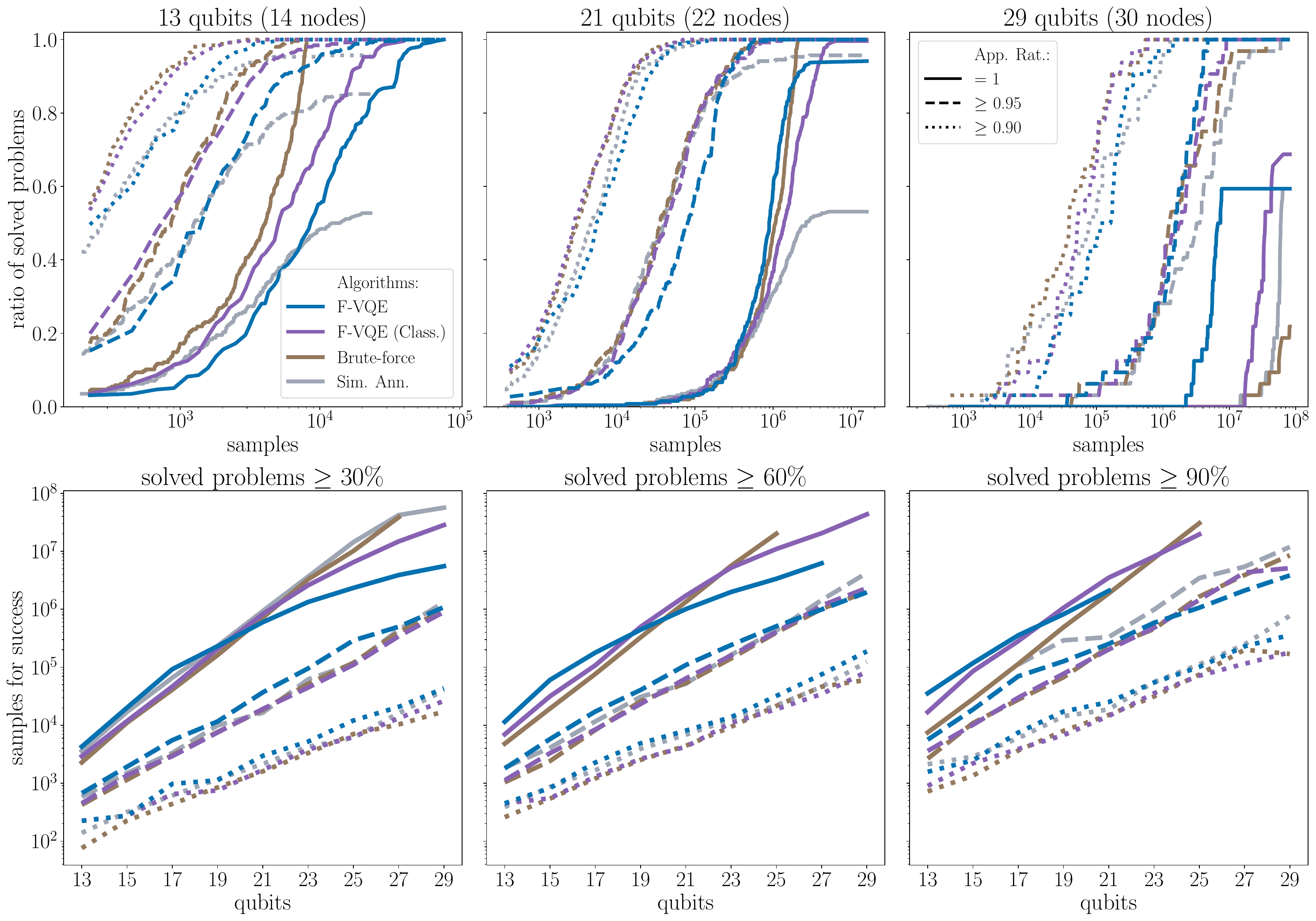}
  \caption{Results of noiseless simulations for weighted MaxCut using F-VQE with the IQP and classical ans\"atze, brute-force search and Simulated Annealing for problem sizes of 13, 15, 17, 19, 21, 23, 25, 27 and 29 qubits. The total number of problems for each qubit size is 256, except for 27 and 29 qubits, where 128 and 32 instances were considered respectively. In the top row, we depict, for the instances of 13, 21, and 29 qubits, the evolution of the percentage of problems that achieve a solution with an approximation ratio larger than 0.9, 0.95 and equal to 1 as a function of the number of samples used along the optimization. Remaining problem sizes can be found in Appendix~\ref{sec:app-sim}. In the bottom row, we present the minimum number of samples for success, i.e. the minimum number required to achieve approximation ratios 0.9, 0.95 and 1 on a fraction of problems 0.3, 0.6 and 0.9, out of the total simulated. If for a given size and method there is no point it means that that fraction for the approximation ratio is not achieved.}
  \label{fig:final_analysis_maxcut}
\end{figure*}
\begin{figure*}[ht!]
  \centering
  \includegraphics[width=2\columnwidth]{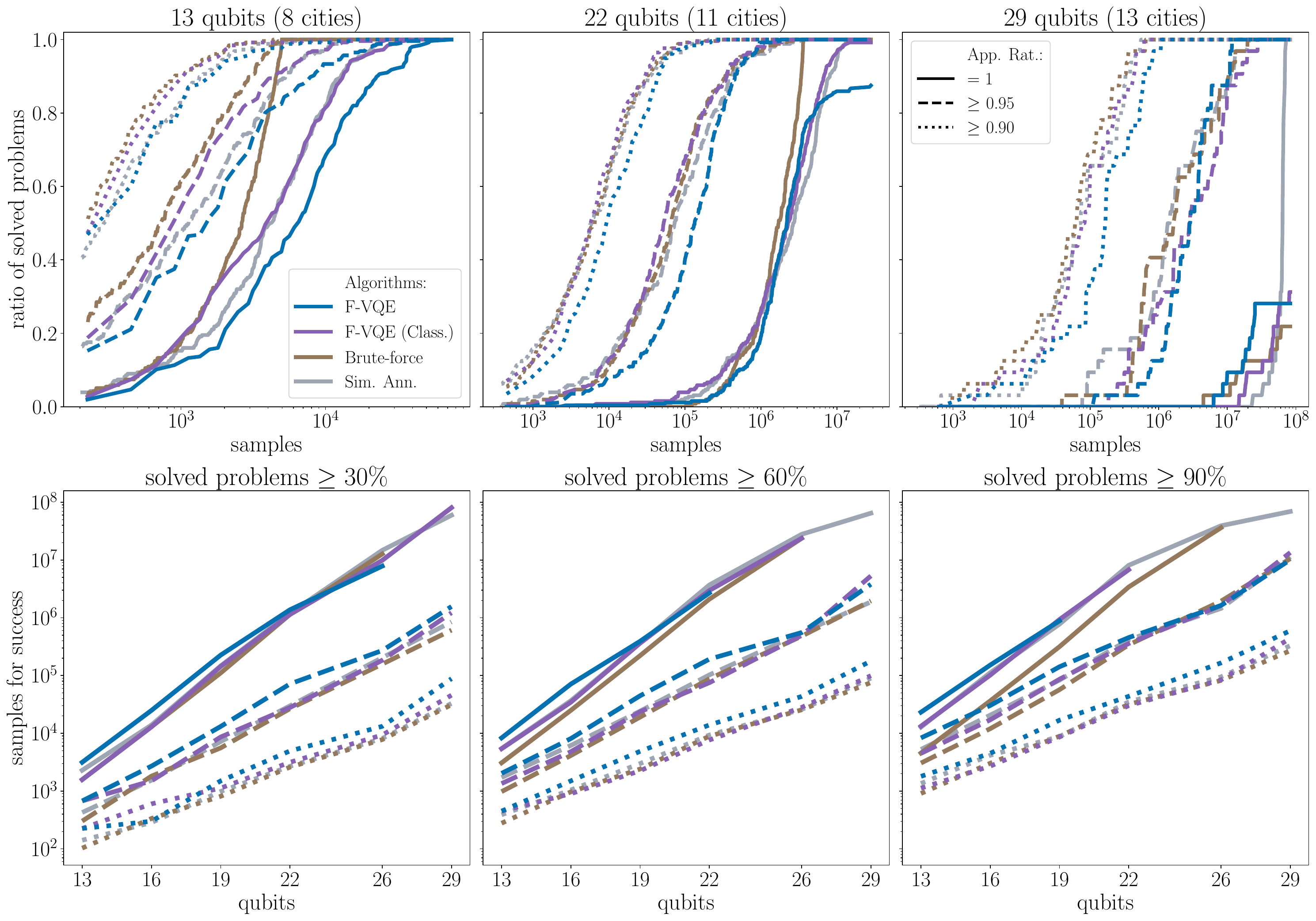}
  \caption{Results of noiseless simulations for ATSP using F-VQE with the IQP and classical ans\"atze, brute-force search and Simulated Annealing for problem sizes of 13, 16, 19, 22, 26 and 29 qubits. The total number of problems for each qubit size is 256, except for 29 qubits, where 32 instances were considered. In the top row, we depict, for the instances of 13, 21, and 29 qubits, the evolution of the percentage of problems that achieve a solution with an approximation ratio larger than 0.9, 0.95 and equal to 1 as a function of the number of samples used along the optimization. Remaining problem sizes can be found in Appendix~\ref{sec:app-sim}. In the bottom row, we have the minimum number of samples for success, i.e. the minimum number of samples required to achieve approximation ratios 0.9, 0.95 and 1 for a fraction of problems 0.3, 0.6 and 0.9, out of the total simulated. If for a given size and method there is no point it means that that fraction for the approximation ratio is not achieved.}
  \label{fig:final_analysis_atsp}
\end{figure*}

\begin{figure*}
  \centering
  \includegraphics[width=1.9\columnwidth]{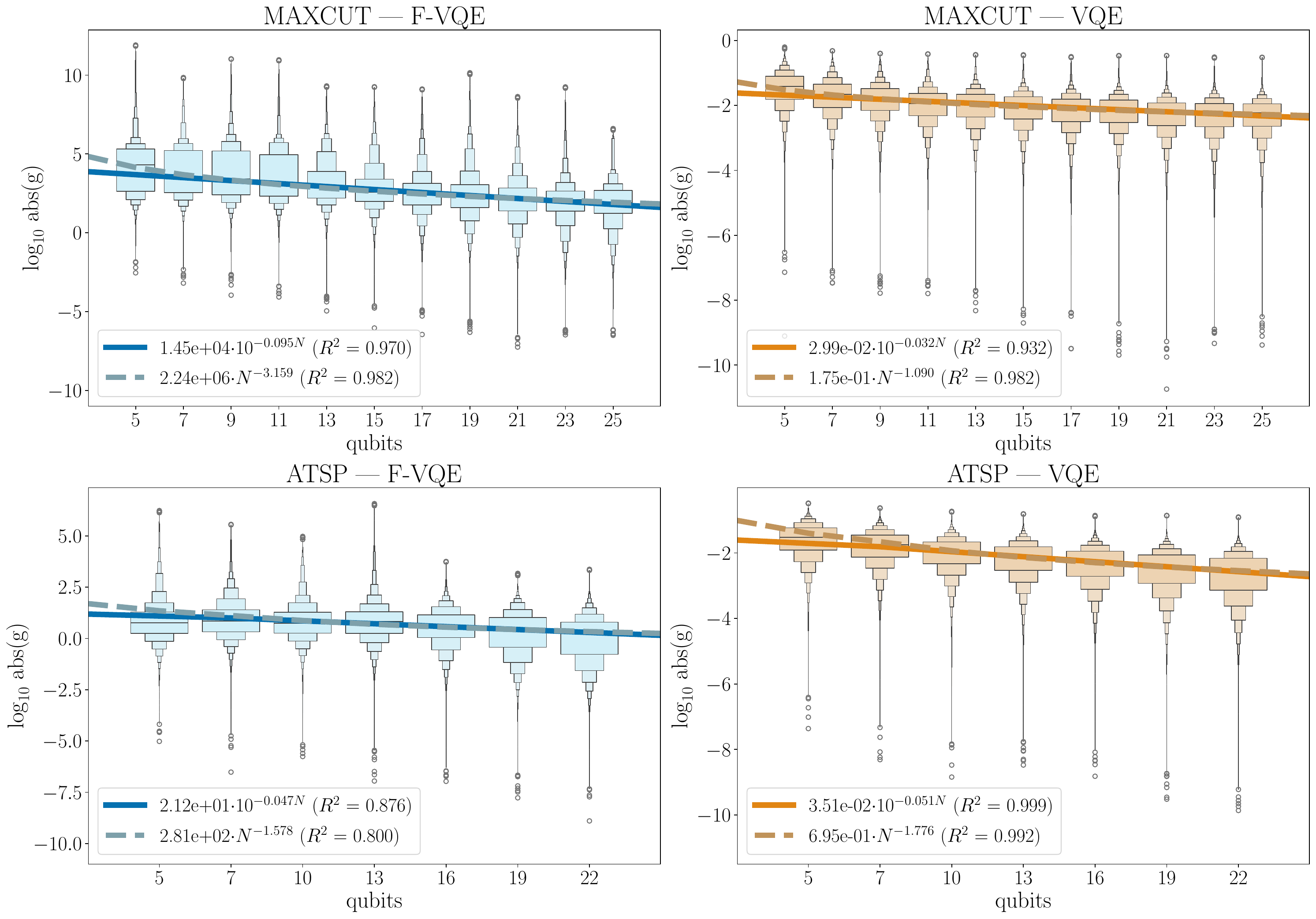}
  \caption{Absolute value of all components of the exact gradient for the 256 simulations per size for both F-VQE (blue) and VQE (orange) for both problem classes: weighted MaxCut (top) and ATSP (bottom). Included in the legends are the regressions for the medians for both exponential and polynomial fits and their $R^2$, where $N$ corresponds to the number of qubits. The best fits have been obtained using data from 7 qubits for MaxCut and from 10 qubits for ATSP.}
  \label{fig:grads}
\end{figure*}

\begin{figure*}
  \centering
  \includegraphics[width=2\columnwidth]{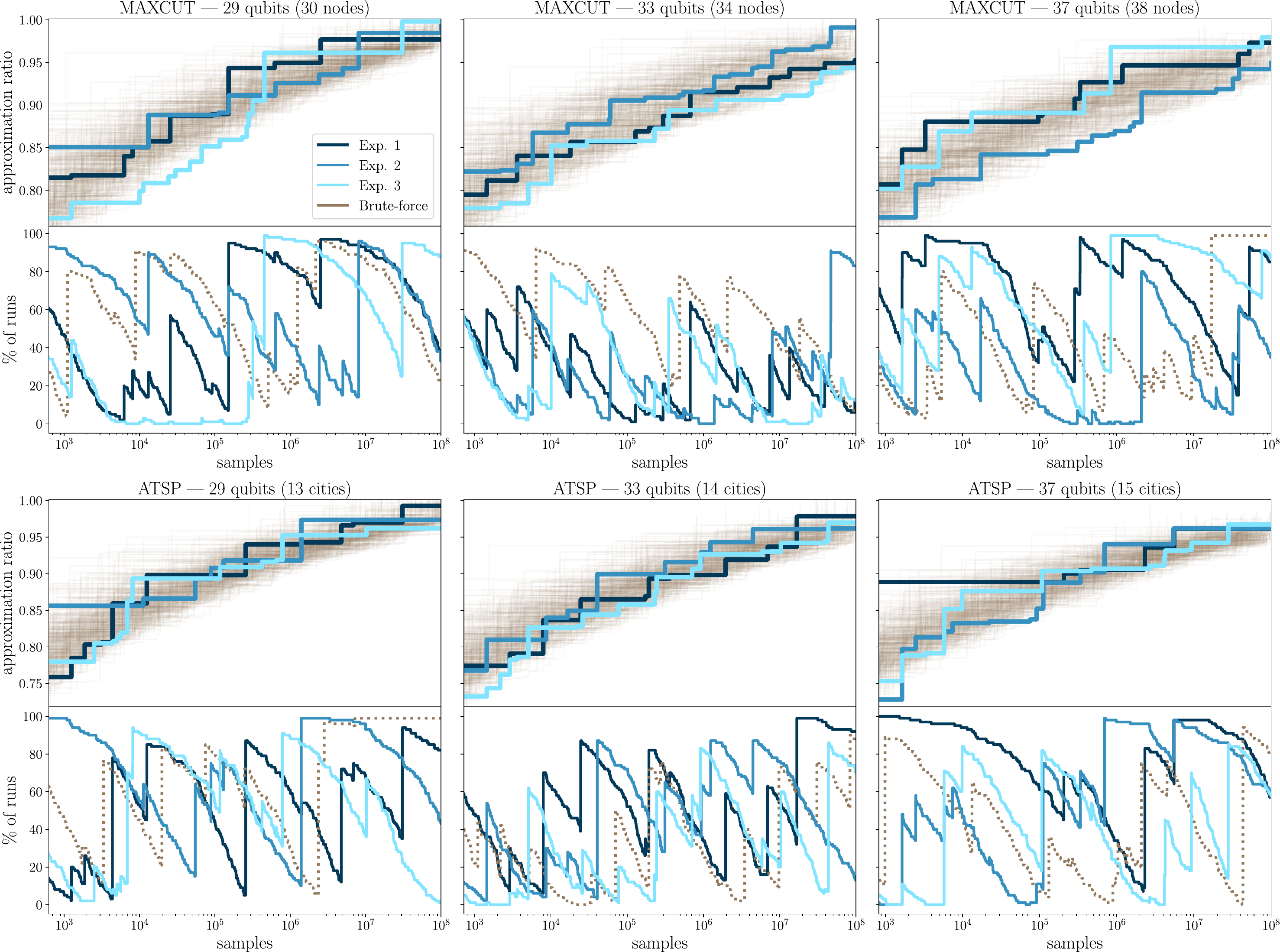}
  \caption{On the upper plots, evolution of the best approximation ratio of the experiments on IBMQ devices for three random instances of weighted MaxCut (top) and ATSP (bottom) problems for 29, 33 and 37 qubits. The solid lines correspond to the experiments listed in Table~\ref{table:ibmq}. The lines in the background correspond to 100 random runs of brute-force search for each of the three problems. Furthermore, on the lower plots, for each size we include the evolution of the percentage of brute-force runs that deliver an equal or worse approximation ratio than their corresponding IBMQ experiment, as well as one of the brute-force runs. Remaining sizes can be found in Appendix~\ref{sec:app-ibmq}.}
  \label{fig:ibmq}
\end{figure*}

\subsection{Noiseless Simulations}\label{sec:sim}
In this experiment we run noiseless simulations of F-VQE with the IQP ansatz (or F-VQE for short) and compare the performance against F-VQE with the classical ansatz, brute-force search, and simulated annealing.

In the top row of Fig.~\ref{fig:final_analysis_maxcut} we illustrate the ratio of MaxCut problems of 13, 21, and 29 qubits for which a solution with an approximation ratio larger or equal than 0.9, 0.95 and the maximum of 1 is achieved. The horizontal axis depicts the number of samples consumed during the execution of the algorithms. This row shows how fast are algorithms capable of solving the considered problem instances. Results for the rest of the problem sizes can be found in Fig.~\ref{fig:final_analysis_maxcut_all} in Appendix~\ref{sec:app-sim}. In the bottom row of Fig.~\ref{fig:final_analysis_maxcut} we show, as a function of the problem size, the minimum number of samples required for every algorithm to achieve a certain approximation ratio on a given percentage of the problems. This row exhibits how rapidly does the consumption of samples grow with the problem size.

In the top row we observe that F-VQE uses less than $8\times10^6$ samples to optimally solve around 60\% of the problems of 29 qubits. The next best algorithm, F-VQE with the classical ansatz, requires over $10^7$ samples to solve its first instance. On the bottom left and middle plots we see that beyond 21 qubits F-VQE needs much less samples than the other algorithms to solve between 30\% and 60\% of problem instances. In those plots F-VQE exhibits a slower exponential growth than BFS, tending even to a sub-exponential trend. 

However, for lower approximation ratios, smaller problems, or when aiming at solving a large fraction of problem instances, F-VQE performs comparably or worse than some of the classical algorithms. In particular, even when given more shots, F-VQE is incapable of increasing the number of solved problem instances, probably due to local minima. As one can see in the top row and in Fig.~\ref{fig:final_analysis_maxcut_all} this maximum ratio of problem instances solved optimally by F-VQE decreases rapidly with the problem size. In contrast, F-VQE with the classical ansatz, once given enough samples, is able to solve more problem instances than F-VQE with the IQP ansatz. Lastly, SA performs poorly across all metrics. Specially, it is not able to find the optimal solution of most problem instances. Probably because our simple version of SA gets quickly stuck in local minima. 
\\

The same analysis is performed for ATSP, and the results are shown in Fig.~\ref{fig:final_analysis_atsp} for 13, 22, and 29 qubits. The rest of problem sizes are depicted in Fig.~\ref{fig:final_analysis_atsp_all}. We observe that both, F-VQE and classical F-VQE, exhibit performance comparable to or worse than BFS. On the other hand, SA performs better for large problem sizes. For 29 qubits (top), SA can solve all problems optimally if given enough samples, a feat not achieved by the other algorithms. We speculate that these results indicate that F-VQE can not exploit the structure of the generalized cost function, reducing the performance to that of unstructured / brute-force search.

\subsection{Barren plateaus} \label{subsec:barr_plat}
Barren plateaus are a common challenge in VQAs~\cite{bp1, CerveroMartin2023barrenplateausin}. In the presence of barren plateaus gradients vanish exponentially in the number of qubits, requiring an exponential amount of samples from the quantum computer to evaluate them with enough precision. For F-VQE, two components could contribute to the presence of barren plateaus: the linear-depth hardware-efficient ansatz, and the use of global loss functions~\cite{Cerezo2021}, i.e., the filter function and the ATSP encoding as a generalized cost function. On the other hand, IQP circuits are not expected to contribute to the presence of barren plateaus, due to their limited expressibility~\cite{Larocca2022, Ragone2023, Bowles2023}. 
In this section we numerically analyse the decrease of the gradients magnitude with the number of qubits to assess if barren plateaus impose a limitation to the scalability of F-VQE.

For every problem instance we compute the exact gradient at every training step and take the absolute value of every partial derivative. We repeat the same process with VQE, where the existence of barren plateaus has been shown in various studies~\cite{Cerezo2021}. 
In Fig.~\ref{fig:grads} we plot the distribution of all the absolute values of the partial derivatives as a function of the number of qubits. We additionally fit exponential and a polynomial decays to the median of the distribution. This provides the decay of a lower bound to the upper half of the distribution. If this bound decayed polynomially with the number of qubits, the algorithm would not suffer the negative effects of barren plateaus.  
 
We observe that the outliers and every quantile decrease with the problem size. For MaxCut, for both F-VQE and VQE, the polynomial decay fits slightly better than the exponential, as can be seen by the higher $R^2$ values. We take this as an indication that the IQP ansatz contributes to avoid barren plateaus, or at least guides the algorithms through a path in the parameter landscape that is free of barren plateaus. For ATSP the exponential regression fits slightly better than the polynomial, though the quality of the fits for F-VQE is significantly lower. This might be caused by the contribution of the generalized encoding to the presence of barren plateaus. We observe no significant qualitative difference between F-VQE and VQE, indicating no obvious benefit nor detriment from the characteristic filtering function of F-VQE. 

\begin{table}[htb!]
\centering
\caption{IBMQ results for weighted MaxCut (left) and ATSP (right).}\label{table:ibmq}%
\begin{tabular}{|cccc||cccc}
\hline
$N$ & IBMQ device   & A.R.   & $\%$ & $N$ & IBMQ device     & A.R.   & \multicolumn{1}{l|}{$\%$} \\ \hline
    & ibm\_cusco    & 0.9980 & 88   &     & ibm\_cusco      & 0.9928 & \multicolumn{1}{l|}{82}   \\
29  & ibm\_cusco    & 0.9920 & 68   & 29  & ibm\_cusco      & 0.9734 & \multicolumn{1}{l|}{43}   \\
    & ibm\_brisbane & 0.9767 & 38   &     & ibm\_brisbane   & 0.9618 & \multicolumn{1}{l|}{01}   \\ \hline
    & ibm\_brisbane & 1      & 100  &     & ibm\_nazca      & 0.9784 & \multicolumn{1}{l|}{92}   \\
31  & ibm\_brisbane & 0.9856 & 50   & 33  & ibm\_cusco      & 0.9697 & \multicolumn{1}{l|}{70}   \\
    & ibm\_brisbane & 0.9804 & 52   &     & ibm\_sherbrooke & 0.9618 & \multicolumn{1}{l|}{13}   \\ \hline
    & ibm\_brisbane & 0.9909 & 83   &     & ibm\_sherbrooke & 0.9673 & \multicolumn{1}{l|}{58}   \\
33  & ibm\_brisbane & 0.9548 & 18   & 37  & ibm\_sherbrooke & 0.9618 & \multicolumn{1}{l|}{57}   \\
    & ibm\_nazca    & 0.9434 & 12   &     & ibm\_brisbane   & 0.9616 & \multicolumn{1}{l|}{57}   \\ \hline
    & ibm\_nazca    & 0.9795 & 87   &     &                 &        &                           \\
37  & ibm\_brisbane & 0.9731 & 85   &     &                 &        &                           \\
    & ibm\_nazca    & 0.9495 & 55   &     &                 &        &                           \\ \cline{1-4}
\end{tabular}
\end{table}
\subsection{IBMQ experiments} \label{sec:ibmq}
Finally, we test the performance of F-VQE with an IQP ansatz in the quantum devices provided by IBMQ. No error correction or mitigation is considered for these experiments. 
We set the maximum number of total samples to $10^8$. To benchmark the results we perform 100 random runs of BFS for each problem instance. In Table~\ref{table:ibmq} we list the devices used for each experiment and the exact approximation ratio of the best solution sampled, up to 4 decimals, for $N\ge29$. Smaller problems, with less candidate solutions than the number of samples we set, have been omitted from the table since their final approximation ratios trivially reach 1. In addition, we include the percentage of BFS runs that obtain an equal or worse final approximation ratio than their corresponding IBMQ experiment.

In Fig.~\ref{fig:ibmq} we show the evolution of the approximation ratio for the weighted MaxCut (top) and ATSP (bottom) for the instances of 29, 33 and 37 qubits. The results for the rest of the problem sizes can be found in Fig.~\ref{fig:ibmq-app} in Appendix~\ref{sec:app-ibmq}. Along with the thick solid lines that correspond to the three IBMQ experiments, we depict the evolution of the approximation ratios for the BFS runs in dotted lines in the background. On the bottom of each figure we plot the evolution of the percentage of BFS runs outperformed by the F-VQE runs and one of the BFS runs chosen randomly. 

We first observe a clear overlap between the evolution of F-VQE and BFS in the upper half of the plots, showing that both algorithms perform similarly. Otherwise, the lines for F-VQE would be consistently above the BFS lines. Plotting the percentage of BFS runs outperformed in the lower half provides more detail. We confirm that F-VQE does not consistently outperform a significant fraction of the BFS runs. Indeed, the random BFS run line behaves similarly to the F-VQE lines. In contrast, Table~\ref{table:ibmq} show that when the entire budget of samples is consumed, F-VQE solves most problems better than 50\% of the BFS runs. The behavior for MaxCut is analogous to the behavior of ATSP.

From these results we conclude that during most of the execution of the algorithm, the performance of F-VQE is comparable to the performance of BFS. We do not consider this sign sufficient to expect that, in the presence of noise, F-VQE will scale better to larger problem sizes.


\section{Conclusions and outlook} \label{sec:conclusions}
In this work we expanded the F-VQE algorithm with the IQP ansatz. This family of circuits is hard to simulate classically, it is not expected to induce barren plateau issues, and as we showed in this work, it can be implemented with a hardware-efficient circuit and halves the number of quantum circuit executions required to estimate gradients of the circuit parameters. 
We then conducted an extensive performance comparison against three classical baseline algorithms. One of them trains a classical ansatz closely related to the IQP ansatz with the F-VQE algorithm to asses the relevance of quantum effects.
We performed noiseless simulations of all the algorithms on weighted MaxCut and ATSP instances of up to 29 qubits and noisy experiments of F-VQE of up to 37 qubits on IBMQ devices. Instead of the usual QUBO form, we employed a natural generalized cost function to encode ATSP with less qubits that does not require introducing additional constraints. The evolution of the approximation ratio and the fraction of problems solved with the number of samples was studied. From the noiseless simulations we additionally studied the effect of barren plateaus. 

We observed some positive signs of the performance of F-VQE. First, it is the algorithm that optimally solves a fraction of the largest simulated MaxCut problems with the least number of samples. More importantly, the scaling of this number with the problem size is better than for any of the classical baseline algorithms. Second, the magnitude of the gradients for MaxCut does not seem to decrease exponentially, unlike for other VQAs subject to barren plateaus. Third, in the IBMQ experiments for a large number of samples F-VQE beats most BFS runs. 

However, we observe negative signs of performance in the remaining regimes of lower approximation ratios, low number of samples, to solve a large ratio of problem instances, or in the presence of noise. Besides, the use of the generalized cost function for ATSP generated barren plateaus and reduced the performance to that of BFS, despite reducing the number of qubits and avoiding the need of penalty terms, compared to the traditional QUBO form. Moreover, we do not consider the difference in performance between the IQP and the classical ansatz significant enough to conclude that the quantum effects in the IQP ansatz play an important role in the performance of F-VQE. 

Looking ahead, under more realistic conditions for solving optimization problems of the relevant scale for industry, the positive performance of F-VQE might disappear. First, we observe that the fraction of problem instances where F-VQE outperforms the classical algorithms decays with the problem size. Second, in the presence of noise the performance of F-VQE drops to the performance of BFS during most of the algorithms execution. Third, for real applications, a more relevant metric than the number of samples is the wall-clock time to execute the algorithm. Since classical operations, like a transition in SA, are significantly faster than running a quantum circuit, F-VQE can become significantly slower than classical algorithms, despite requiring fewer samples. 
Fourth, the binary encoding that F-VQE and other VQAs require to formulate optimization problems might reduce their performance on some problem classes.
Finally, for a practical quantum advantage, F-VQE would need to outperform the state-of-the-art classical algorithms such as Gurobi~\cite{gurobi} and CPLEX~\cite{cplex2009v12} at least marginally or in some relevant metric like the runtime, the solution quality, or the economic cost of their implementation.

Under these considerations we do not expect F-VQE to provide an advantage over classical methods in realistic conditions without significant algorithmic and hardware development.

\bibliography{bibliography}


\clearpage

\onecolumngrid

\appendix
\section{Gradients}\label{sec:app-grads}
\subsection{F-VQE}
In this section we derive Eq.~(\ref{eq:grad-prob}). Let us first compute the partial derivative from the loss function defined for F-VQE in Eq.~(\ref{eq:loss}):
\begin{align}
    \dfrac{\partial \mathcal{L}_t(\bm{\theta})}{\partial \theta_k}&=-\dfrac{\partial\left(\sum_{\bm{x}\in\{0,1\}^N} \sqrt{P_{\psi(\bm{\theta})}(\bm{x})P^{(f)}_{\psi_{t-1}}(\bm{x})}\right)^2}{\partial \theta_k}\\
    &=-\left(\sum_{\bm{x}\in\{0,1\}^N} \sqrt{P_{\psi(\bm{\theta})}(\bm{x})P^{(f)}_{\psi_{t-1}}(\bm{x})}\right)
    \cdot\sum_{\bm{x}\in\{0,1\}^N}\frac{P^{(f)}_{\psi_{t-1}}(\bm{x})}{\sqrt{P_{\psi(\bm{\theta})}(\bm{x})P^{(f)}_{\psi_{t-1}}(\bm{x})}}\dfrac{\partial P_{\psi(\bm{\theta})}(\bm{x})}{\partial \theta_k}.
\end{align}

Now, let us apply this partial derivative on the last step $t-1$:
\begin{align}
    \left. \dfrac{\partial \mathcal{L}_t(\bm{\theta})}{\partial \theta_k} \right|_{\bm{\theta}_{t-1}} =& -\left(\sum_{\bm{x}\in\{0,1\}^N} \sqrt{\frac{P^2_{\psi_{t-1}}(\bm{x})f^2(C(\bm{x});\tau)}{\mathbb{E}_{\psi_{t-1}}(f^2)}}\right)
    \cdot \sum_{\bm{x}\in\{0,1\}^N}\sqrt{\frac{f^2(C(\bm{x});\tau)}{\mathbb{E}_{\psi_{t-1}}(f^2)}} \left.\dfrac{\partial P_{\psi(\bm{\theta})}(\bm{x})}{\partial \theta_k}\right|_{\bm{\theta}_{t-1}}\\
    =&-\frac{\mathbb{E}_{\psi_{t-1}}(f)}{\mathbb{E}_{\psi_{t-1}}(f^2)}\sum_{\bm{x}\in\{0,1\}^N}f(C(\bm{x});\tau)\left.\dfrac{\partial P_{\psi(\bm{\theta})}(\bm{x})}{\partial \theta_k}\right|_{\bm{\theta}_{t-1}}.\label{eq:app_grad}
\end{align}
Finally, applying the parameter-shift rule on $P_{\psi(\bm{\theta})}(\bm{x})$ we get that the gradient is, as expected,
\begin{equation}
    \left. \dfrac{\partial \mathcal{L}_t(\bm{\theta})}{\partial \theta_k} \right|_{\bm{\theta}_{t-1}} =\frac{\mathbb{E}_{\psi_{t-1}}(f)}{\mathbb{E}_{\psi_{t-1}}(f^2)}\left(\mathbb{E}_{\psi_{t-1}^{-k}}(f) - \mathbb{E}_{\psi_{t-1}^{+k}}(f)\right).
\end{equation}

\subsection{Classical ansatz}
In this section we derive the gradients for F-VQE with the classical ansatz. Notice that Eq.~\ref{eq:app_grad} is still valid for this new ansatz except that we now have a classical state $\rho(\bm{\theta})$ instead of the state $\psi(\bm{\theta})$. Let us now compute the partial derivative of the classical state. First, since all the components from the classical ansatz commute with each other, given any parameter $\theta_k$ for $k\in[M]$ we can write
\begin{align}
    \rho(\bm{\theta}) &= \mathcal{E}_{\theta_M}\circ\dots\circ\mathcal{E}_{\theta_1} \left(\ket{0}\bra{0}^{\smallotimes  N}\right)
    =  \mathcal{E}_{\theta_k}\circ\dots\mathcal{E}_{\theta_{k+1}}\circ\dots\mathcal{E}_{\theta_{k-1}}\circ\dots\circ\mathcal{E}_{\theta_1} \left(\ket{0}\bra{0}^{\smallotimes  N}\right)\\
    &=\mathcal{E}_{\theta_k}(\rho((\theta_1, ..., \theta_{k-1}, \theta_{k+1}, ..., \theta_{M}))\equiv\mathcal{E}_{\theta_k}(\hat{\rho_{k}}),
\end{align}
where $\hat{\rho_{k}}$ denotes the circuit $\rho(\bm{\theta})$ without the component $k$. Then, the partial derivative is
\begin{align}
    \dfrac{\partial \rho(\bm{\theta})}{\partial \theta_k}&=\dfrac{\partial \mathcal{E}_{\theta_k}(\hat{\rho_{k}})}{\partial \theta_k}
    =-\cos\left(\frac{\theta_k}{2}\right)\sin\left(\frac{\theta_k}{2}\right)\left(\hat{\rho_{k}}-X_{Q_k}\hat{\rho_{k}}X_{Q_k}\right).
\end{align}
Applying the parameter-shift rule as in the quantum case, we find that we obtain the same result by shifting the parameter by an amount $\pi/2$.
\begin{equation}
    \dfrac{\partial \rho(\bm{\theta})}{\partial \theta_k} = \frac{1}{2}\rho(\bm{\theta}-\pi \bm{e}_k /2) - \frac{1}{2}\rho(\bm{\theta}+\pi \bm{e}_k /2).
\end{equation}
As with the IQP ansatz, we can also obtain the probability distribution of the second state from the first distribution of the first state. Finally, if we plug this result into Eq.~\ref{eq:app_grad} we obtain
\begin{align}
    \left. \dfrac{\partial \mathcal{L}_t(\bm{\theta})}{\partial \theta_k} \right|_{\bm{\theta}_{t-1}} =&\frac{\mathbb{E}_{\rho_{t-1}}(f)}{\mathbb{E}_{\rho_{t-1}}(f^2)}\left(\mathbb{E}_{\rho_{t-1}^{-k}}(f) - \mathbb{E}_{\rho_{t-1}^{+k}}(f)\right).
\end{align}

\section{IQP to classical ansatz via dephasing} \label{app:dephasing}
In this appendix we show that the IQP ansatz presented in Fig.~\ref{fig:ansatz} reduces to the classical ansatz of Eqs.~(\ref{eq:class_ans}, \ref{eq:bitflip}) under a completely dephasing channel 
\begin{equation}
    \mathcal{D}_q(\rho) = \frac{1}{2}\rho + \frac{1}{2}Z_q\rho Z_q
\end{equation}
applied on every qubit after every single-qubit rotation $R_x(\theta_k)$. Here $Z_q$ is the Pauli-$Z$ operator acting on qubit $q$. Notice that the channel acts trivially on classical probability distributions over computational states:
\begin{equation}
    \mathcal{D}_q(\rho) = \rho \quad \forall \, \rho = \sum_{\bm{x}} P(\bm{x})\ket{\bm{x}}\bra{\bm{x}}.
\end{equation}

In every layer $\ell$ of the IQP circuit, single-qubit $R_{q\ell} =  \exp(-i \theta_{k} X_q/2)$ rotations are applied on all qubits. We are using a slightly different notation to mark more clearly the qubit where the rotation has support. Labels $k$ have a one-to-one correspondence to the position $(q,\ell)$ of the rotation in the circuit. If these are followed by single-qubit dephasing channels, one can check that their action on the state becomes a probabilistic bit-flip
\begin{equation}
    \mathcal{D}_q\left(R_{q\ell}\rho R_{q\ell}^\dagger\right) = \frac{1}{2}R_{q\ell}\rho R_{q\ell}^\dagger + \frac{1}{2}R_{q\ell}^\dagger\rho R_{q\ell} = \cos^2(\theta_k/2) \rho + \sin^2(\theta_k/2) X_q \rho X_q \equiv \mathcal{E}_{q\ell}(\rho)
\end{equation}
analogous to the ones that constitute the classical ansatz. 

Let's follow each operation in the IQP slowly. As in the classical ansatz, we start on the classical state $ \ket{0}\bra{0}^{\smallotimes N}$. This is followed by a layer of single-qubit rotations and dephasing channels, so that the resulting state is a composition of classical single-qubit bit-flip channels 
\begin{equation}
    \mathcal{E}_{N1}\circ \cdots \circ \mathcal{E}_{11}\left(\ket{0}\bra{0}^{\smallotimes N}\right).
\end{equation}
The column of CNOTs acts on the state by changing the support of the bit-flip channels to the support of the bit-flips in the classical ansatz of Eq.~\eqref{eq:bitflip}:
\begin{equation}
    \mathcal{E}_{q\ell} \rightarrow \mathcal{E}_{\theta_k}.
\end{equation}

Applying the same procedure to every layer shows finally that under the completely dephasing channel the IQP ansatz reduces to the classical ansatz.

\section{Problem generation}\label{sec:app-pg}
\begin{figure}[hbt!]
  \centering
  \includegraphics[width=1.\columnwidth]{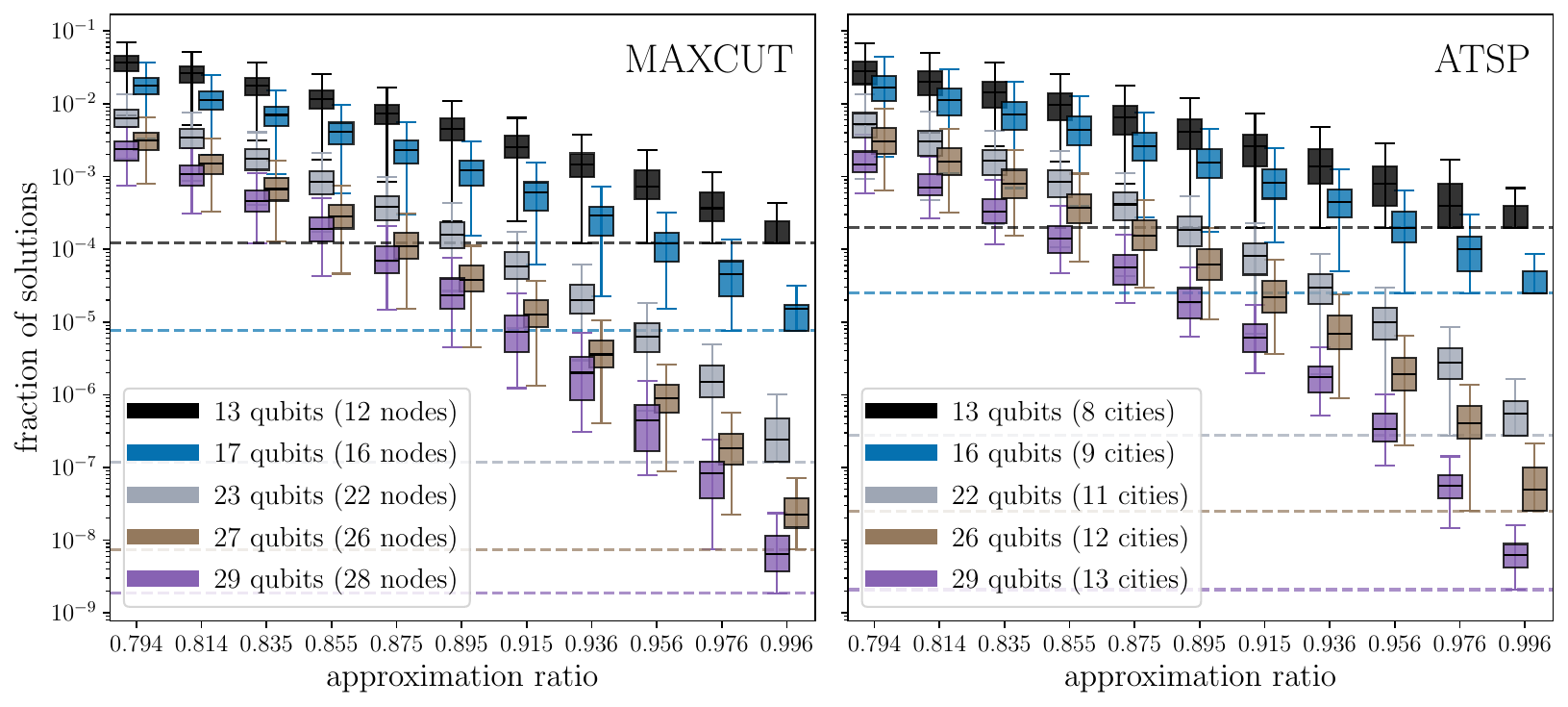}
  \caption{Distributions of the fraction of solutions with an approximation ratio $\geq x$, with $x$ taking values in between 0.794 and 0.996, for all the generated problem instances of different sizes for both weighted MaxCut (left) and ATSP (right). For reference, the horizontal dashed lines indicate the fraction of having only one optimal solution per each size and problem class. For MaxCut this line is exactly $2^{-N}$, while for ATSP is $1/(N_\text{cities}-1)!$.}
  \label{fig:spectrum}
\end{figure}

For this work's simulations we have generated 256 random problem instances of each problem class for qubit sizes of up to 26, 128 for 27 qubits and 32 for the 29 qubits case. In addition, we have also generated 3 instances of qubit sizes of 19, 25, 29, 31, 33 and 37 (19, 22, 26, 29, 33 and 37) for weighted MaxCut (ATSP) for the experiments on real devices. In Fig.~\ref{fig:spectrum} we depict the fraction of solutions with equal or larger approximation ratio than a given value for the instances generated of certain sizes. To compute this, given a set of problems of a type and size, we have counted the number the number of solutions with approximation ratios larger than a certain value and plotted the distribution. Notice that the fraction of solutions with larger approximation ratios is very low, making these problems hard to solve. Include that some problems have degeneracy in the optimal solution

\begin{figure*}[tbp!]
  \centering
  \includegraphics[width=1.\columnwidth]{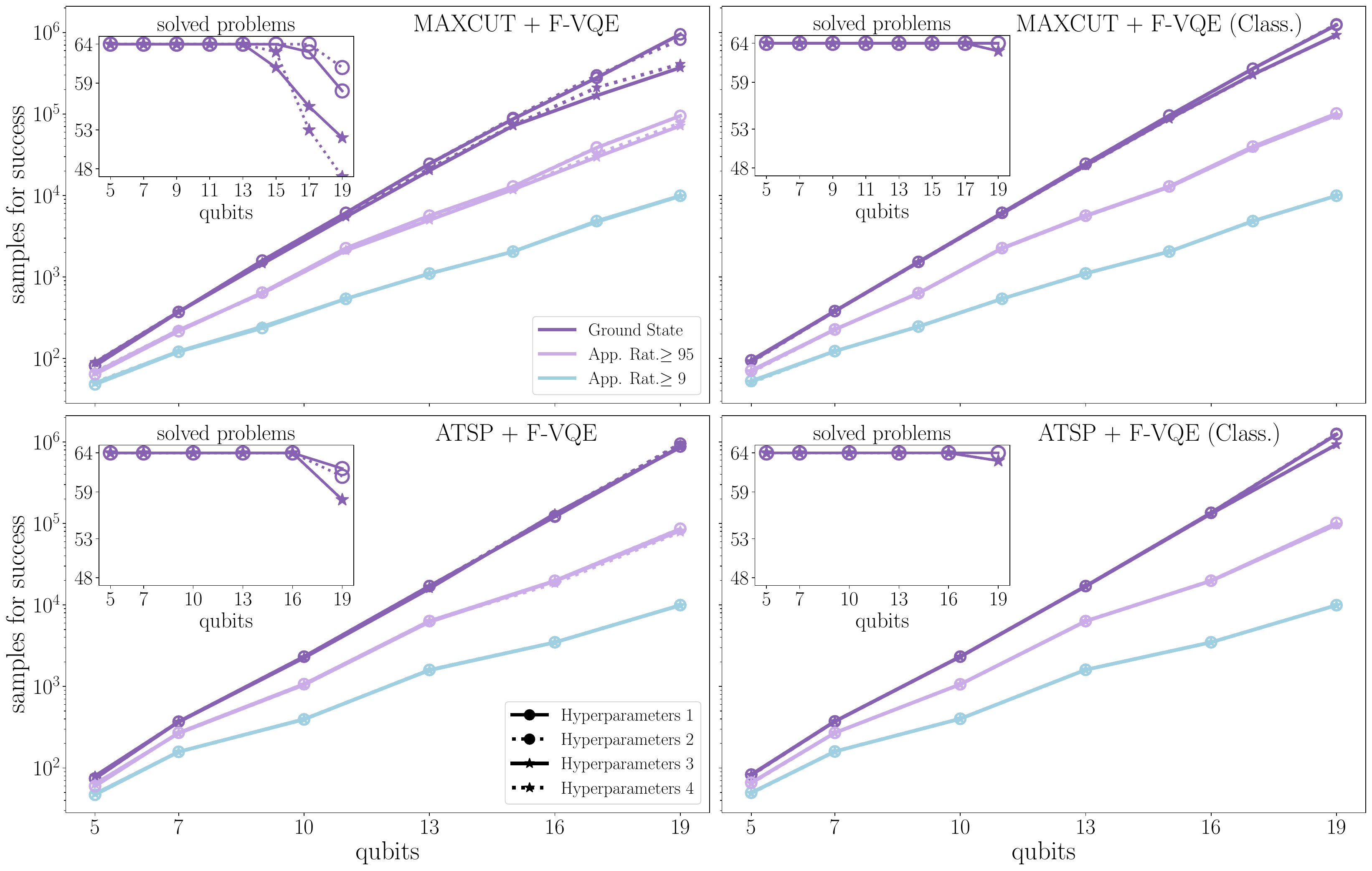}
  \caption{Mean of the number of samples for success, i.e. to obtain a cumulative probability larger than 95\%, of the simulation of 64 random instances of weighted MaxCut (top) and ATSP (bottom) for different qubit sizes and for both algorithms, F-VQE with the IQP (left) and the classical (right) ans\"atze. We have a comparison for the four sets of hyperparameters that can be found in Table~\ref{table:hyperparameters} and for not only the ground state but for two different values of approximation ratio. We also have the number of experiments that leads to such cumulative probability for the optimal solutions, seen in each of the subplots.}
  \label{fig:hyperparameters}
\end{figure*}
\section{Hyperparameters}\label{sec:app-hyp}
Let us now analyze the performance of F-VQE with the IQP and classical ans\"atze using different sets of hyperparameters for both optimization problems: weighted MaxCut and ATSP. We consider four situations: large and small number of shots, for variable and constant filtering strength and learning rate, as depicted in Table~\ref{table:hyperparameters}. 

In Fig.~\ref{fig:hyperparameters} we present the results of the exact simulations of 64 random problem instances of weighted MaxCut and ATSP using the four different sets of hyperparameters for the two types of ans\"atze. Depicted are the mean number of samples required to achieve an exact cumulative probability larger than 95\% for obtaining solutions of different approximation ratio values (1, 0.95 and 0.9). In addition, we include a subplot with the total number of experiments that lead to such cumulative probability for the optimal solutions. Observe that for both large and small number of shots, drawn as circles and stars respectively, the behavior of using constant and variable hyperparameters, continuous and dashed lines, is very similar. We do see a difference on the number of samples for success, where the sets with less shots require less samples. However, the number of solved problems decreases faster for these sets than the ones using more shots. Given these results, we decide to use the set of Hyperparameters 2 (large number of shots and constant $\tau$ and $\eta$) for both optimization problems and ans\"atze.

\begin{table}
\caption{Hyperparameters considered, with $N$ the number of qubits.}
\centering
\begin{tabular}{c|ccc}
                  & \# of shots                         & $\tau$                   & $\eta$                           \\ \hline
HP. 1 & $25 \times N - 100$                 & $1 + 0.1 \times N$       & $0.45 - 0.01 \times N$         \\
HP. 2 & $25 \times N - 100$                 & $2.5$                    & $0.25$                         \\
HP. 3 & $ 2.5 \times N - 10 $ & $(1 + 0.1 \times N) / 5$ & $(0.45 - 0.01 \times N) / 1.5$ \\
HP. 4 & $ 2.5 \times N - 10 $ & $0.25$                   & $0.125$                       
\end{tabular}
\label{table:hyperparameters}%
\end{table}


\section{Noiseless Simulations}\label{sec:app-sim}
\begin{figure*}
  \centering
  \includegraphics[width=1\columnwidth]{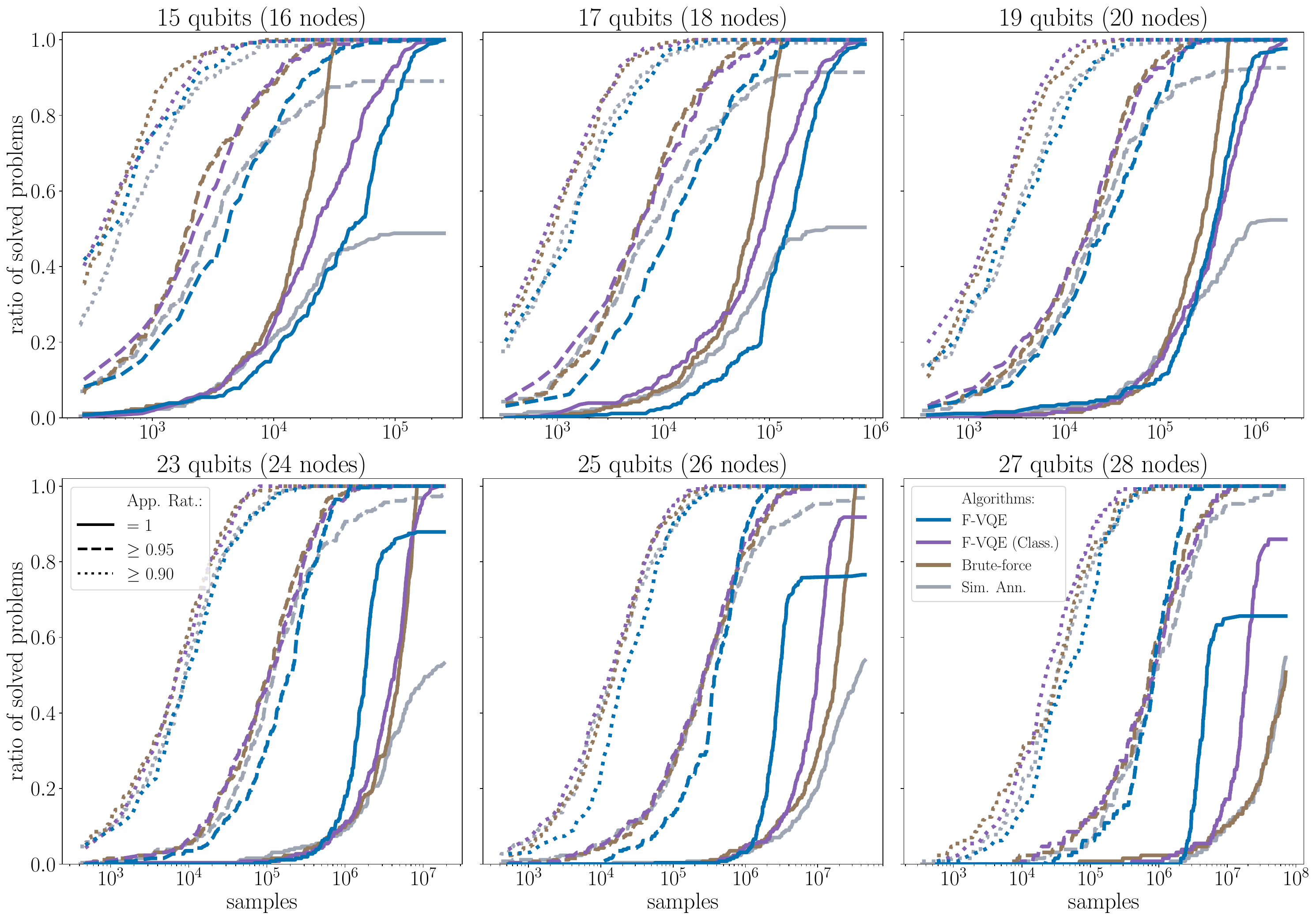}
  \caption{Complementary figure to Fig.~\ref{fig:final_analysis_maxcut}. Results of noiseless simulations for weighted MaxCut using F-VQE with the IQP and classical ans\"atze, brute-force search and Simulated Annealing for problem sizes of 15, 17, 19, 23, 25 and 27 qubits. The total number of problems for each qubit size is 256, except for 27 qubits, where 128 instances were considered. We depict the evolution of the percentage of problems that achieve a solution with an approximation ratio larger than 0.9, 0.95 and equal to 1 as a function of the number of samples used along the optimization for the instances of each problem size.}
  \label{fig:final_analysis_maxcut_all}
\end{figure*}

\begin{figure*}[ht!]
  \centering
  \includegraphics[width=1\columnwidth]{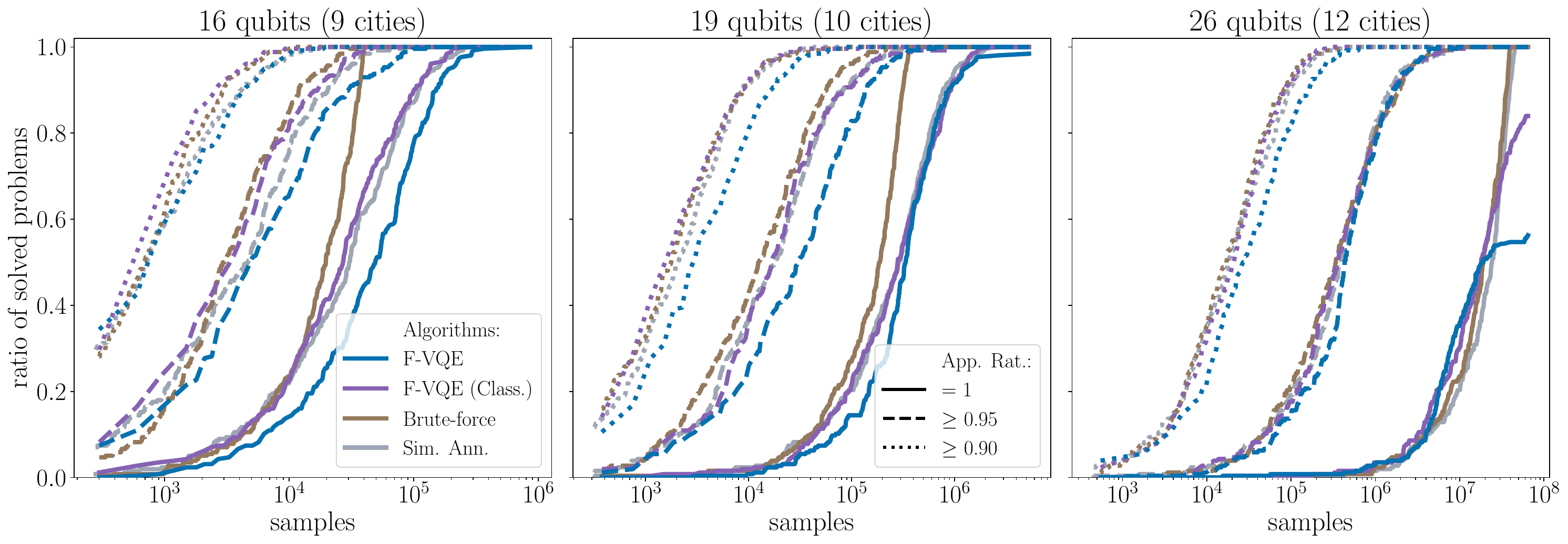}
  \caption{Complementary figure to Fig.~\ref{fig:final_analysis_atsp}. Results of noiseless simulations for ATSP using F-VQE with the IQP and classical ans\"atze, brute-force search and Simulated Annealing for problem sizes of 16, 19 and 26 qubits. The total number of problems for each qubit size is 256. We depict the evolution of the percentage of problems that achieve a solution with an approximation ratio larger than 0.9, 0.95 and equal to 1 as a function of the number of samples used along the optimization for the instances of each problem size.}
  \label{fig:final_analysis_atsp_all}
\end{figure*}

Complementary results to Section~\ref{sec:sim}. In Fig.~\ref{fig:final_analysis_maxcut_all} we depict the results of noiseless simulations for weighted MaxCut using F-VQE with the IQP and classical ans\"atze, BFS and SA for problem sizes of 15, 17, 19, 23, 25 and 27 qubits. Finally, in Fig.~\ref{fig:final_analysis_atsp_all} we show the results of the same study for ATSP for sizes of 16, 19 and 26 qubits.

\section{IBMQ Experiments}\label{sec:app-ibmq}
\begin{figure*}
  \centering
  \includegraphics[width=1\columnwidth]{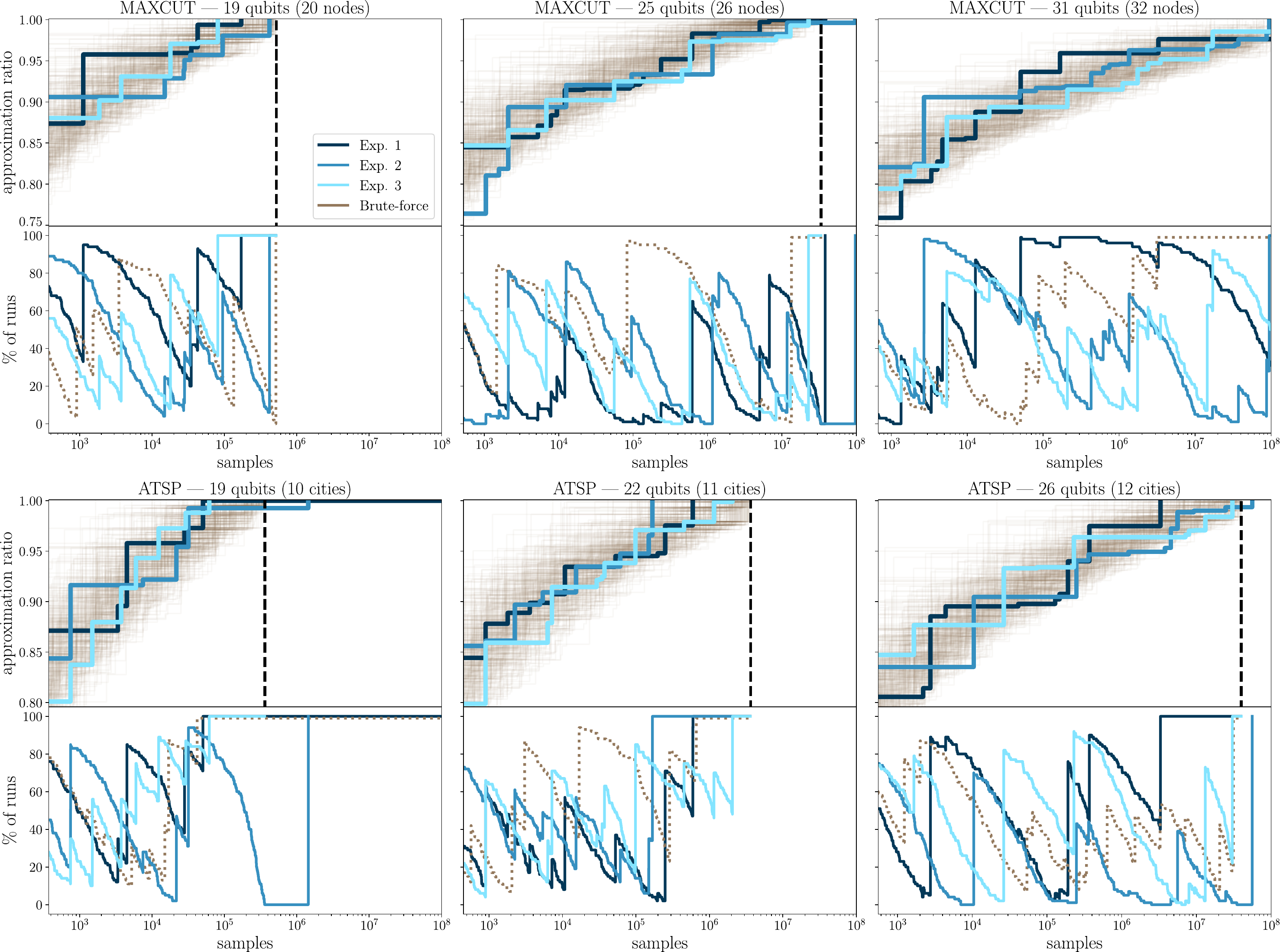}
  \caption{Complementary figure to Fig.~\ref{fig:ibmq}. On the upper plots, evolution of the best approximation ratio of the experiments on IBMQ devices for three random instances of 19, 25 and 25 qubits for weighted MaxCut (top) and 19, 22 and 26 qubits for ATSP. The solid lines correspond to the experiments listed in Table~\ref{table:ibmq}. The lines in the background correspond to 100 random runs of brute-force search for each of the three problems. Furthermore, on the lower plots, for each size we include the evolution of the percentage of brute-force runs that deliver an equal or worse approximation ratio than their corresponding IBMQ experiment, as well as one of the brute-force runs. Finally, we also include the total number of solutions, $2^N$ for MaxCut and $(N_\text{cities}-1)!$ for ATSP, as a vertical gray dashed line as long as it is less than the maximum total samples of $10^8$, being $N$ the number of qubits.}
  \label{fig:ibmq-app}
\end{figure*}

\begin{figure*}
  \centering
  \includegraphics[width=1\columnwidth]{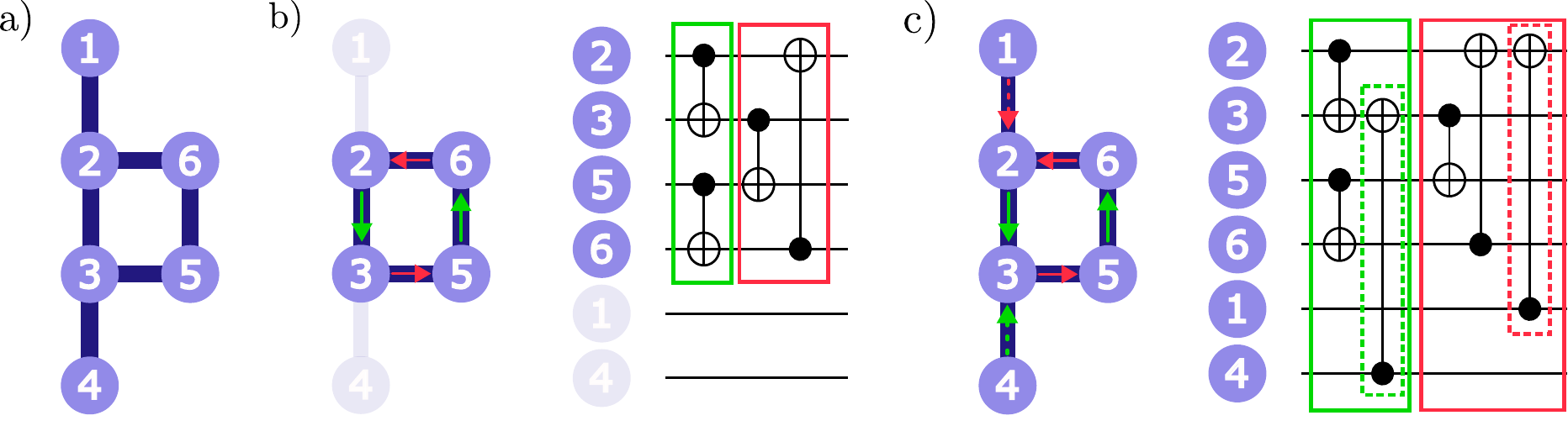}
  \caption{Example of how to adapt the IQP ansatz to a 6-qubit device when it is not possible to find a sufficiently long chain of qubits. First, in a) we show the qubit connectivity of the device. Note that it is possible to connect 5 qubits in a chain, but not 6, so using the same CNOT structure presented in Fig.~\ref{fig:ansatz} would require SWAP gates. To avoid this, we first select the longest cycle in the device, which in this case comprises qubits 2, 3, 5 and 6, as shown in b). The arrows represent the CNOT gates, pointing from the control to the target. Also, color green (red) indicates that the CNOT corresponds to the first (second) column in the circuit. Given the cycle 2-3-5-6-2, we assign to each connection an arrow and alternating color, starting with green. Consequently, in the first column we have CNOTs with controls in 2 and 5 and then 3 and 6 being the corresponding targets. Similarly, in the second column we have the other CNOTs with controls (targets) in 3 and 6 (5 and 2). Finally, in c) we integrate the remaining qubits into the circuit by connecting them to the cycle. To do so, we assign arrows of the same color that enters the connecting qubit. In the example, we assign a red (green) arrow from qubit 1 to 2 (4 to 3). This corresponds to adding a CNOT with control in 1 (4) and target in 2 (3) in the second (first) column of the circuit.}
  \label{fig:cnot}
\end{figure*}

Complementary results to Section~\ref{sec:ibmq}. In Fig.~\ref{fig:ibmq-app} we expand the results presented in Fig.~\ref{fig:ibmq} to the remaining sizes of 19, 25 and 31 qubits for weighted MaxCut and 19, 22 and 26 qubits for ATSP. In Fig.~\ref{fig:cnot} we show an example of how we adapt the CNOT configuration of the IQP ansatz to a given device when it is not possible to find a sufficiently long chain of qubits. In our experiments, this procedure is applied to the problem sizes with $N=22, 25$ and 26 .

\end{document}